\documentclass{article}
\usepackage[utf8]{inputenc}
\usepackage{graphics}
\usepackage{bm}
\usepackage{graphicx}
\usepackage{amsmath}
\usepackage{amssymb}
\usepackage{amstext}
\usepackage{amscd}
\usepackage{amsfonts}
\usepackage{epstopdf}
\usepackage{color}
\usepackage{hyperref}
\title{Two approaches that prove divergence free nature of non-local gravity}

\author{{M. Hameeda$^{1,2,a}$, B. Pourhassan$^{3,b}$, M.C.Rocca$^{4,5,6,c,d}$, }\\
{\texttt{ \rm{ Aram Bahroz Brzo$^{7,e}$}}}\\
\small{$^1$ Department of Physics, S.P. College, Srinagar, Kashmir, 190001 India}\\
\small{$^2$ Inter University Centre for Astronomy and Astrophysics , Pune India}\\
\small{$^3$ School of Physics, Damghan University,}\\
\small{ P. O. Box 3671641167, Damghan, Iran}\\
\small{$^4$ Departamento de F\'{\i}sica,
Universidad Nacional de La Plata,}\\
\small{$^5$ Departamento de Matem\'{a}tica,
Universidad Nacional de La Plata,}\\
\small{$^6$ Consejo Nacional de Investigaciones Cient\'{\i}ficas
y Tecnol\'{o}gicas}\\
\small{(IFLP-CCT-CONICET)-C. C. 727, 1900 La Plata -
Argentina}\\
\small{$^7$ Department of Physics, College of Education,}\\
\small{ University of Sulaimani, Kurdistan-Iraq}\\
\small{\texttt{\rm{$^{a}$hme123eda@gmail.com, $^{b}$b.pourhassan@du.ac.ir,}}}\\
\small{\texttt{\rm{$^{c}$rocca@fisica.unlp.edu.ar,$^{d}$mariocarlosrocca@gmail.com, }}}\\
\small{\texttt{\rm{$^{e}$aram.brzo@univsul.edu.iq }}}}
\date{\today}

\begin{document}

\maketitle

\begin{abstract}
This paper is an attempt to study the thermodynamics of the structure formation in the large scale universe in the non local gravity using Boltzmann statistics and the Tsallis statistics. The partition function is obtained in both the approaches and the corresponding thermodynamics properties are evaluated. The important thing about the paper is that we surprisingly get the divergence free integrals and thus stress upon the fact that the nonlocal gravity is the singularity free model of gravity.
\end{abstract}

\maketitle

\section{Introduction}

The Einstein's interpretation of the well known principle of equivalence of inertial and gravitational masses in terms of the proximate relation between inertia and gravitation \cite{ein} which ensuingly led to the establishment of the extremely local principle of equivalence and GR. Following Einstein, a general connection between inertia and gravitation has been employed as a guiding principle to render GR nonlocal in just the same manner that accelerated observers in Minkowski spacetime are nonlocal.\\ 
The general problems with action principles for nonlocal theories in regard to the  symmetric issues of kernel has been elaborated in Hehl and Mashhoon (2009b) and a classical nonlocal generalization of Einstein's theory of gravitation has been extensively developed \cite{bah,bah1,bah2,bah3}. The emergence of  nonlocality for instance from integrating out certain physical degrees of freedom has also been discussed \cite{gal}.\\
Nonlocal gravity (NLG) is originally a tetrad theory established upon the frame field of a fundamental family of observers in spacetime with the gravitational field characterized by torsion, which is most directly related to the tetrad frame field of the fundamental observers. The form of the field equation of nonlocal gravity gives rise to the possibility that nonlocal gravity may simulate dark matter \cite{mas}.\\
The classical action of gravity being local and is usually used to develop the cosmological models be it the Einstein-Hilbert action coupled to matter or some modified gravity theory, the non locality effect comes from the quantum fluctuations and gets incorporated into the corresponding quantum effective action.  Well understood in the ultraviolet regime but much less in the infrared, these nonlocalities could in principle give rise to important cosmological effects. \cite{eni}.  The main difference between the classical action which is fundamentally local and the quantum effective action is that, the signs of nonlocality starts surfacing whenever the theory contains massless or light particles.\\
The essential difference between nonlocal gravity and general relativity has been expressed in terms of spacetime memory or the presence of effective dark matter. In terms of the memory from the past, nonlacality has been considered a natural feature of the universal gravitational interaction. Some of the consequences of non local gravity for instance the nonlocal modifications of Newtonian gravity and linearized gravitational waves have been given a detailed attention \cite{non}.\\
The isotropy of the cosmic microwave background radiation is a strong indication that a small amplitude inhomogeneities must have existed at the time of decoupling. The intrinsic gravitational instability of a nearly homogeneous distribution of matter has lead to the colossal growth of such inhomogeneities from the recombination era to the present.\\
The exact manner in which structure formation like the formation of galaxies, clusters of galaxies and thereafter the spread of the cosmic web have come about is still a mystery. But a general belief is that dark matter has played a crucial role in this whole development.
This has been known and studied that under the assumptions of spatial homogeneity and isotropy, and as long as the
net pressure, as being assumed as a source of gravity, can be neglected in comparison with the energy density associated with the matter content of the universe,  Newtonian cosmology is an ultimate approximation to the standard Friedmann-Lemaitre-Robertson-Walker (FLRW) cosmological models of general relativity.  More generally, it has been sort out that after recombination era, Newtonian gravitation can be efficiently applied in the study of nonrelativistic motion of matter on subhorizon scales \cite{pee,zel,muk,gur}.
The important task remains is to investigate whether
nonlocal gravity is fullfilling the capability of solving the problem of large-scale cosmological structure formation.\\
The clustering of galaxies, their distribution functions and the corresponding thermodynamics  have been extensively studied using Newtonian gravity both for point masses and for extended masses \cite{fmh,ahm06,sas,sas2,ahm10,sas84,Hameeda2}. The modification in the gravity action can lead to the variations in  the gravitational potential in the low-energy limit and the modified potential reduces to the Newtonian one on the Solar system scale as well. It has been analyzed in \cite{mil1,mil2,mil3} that the modified gravitational potential could fit galaxy rotation curves even without considering dark matter or dark energy. This in fact has, provided  opportunities to draw a formal analogy between the corrections due to the modified Newtonian potential and the dark energy models.\\
The clustering of galaxies has been studied by considering Newtonian modified potential of a brane world model by using standard techniques of statistical mechanics and the thermodynamics of gravitational clustering of galaxies have been discussed \cite{Hameeda}. The modification to the Newtonian potential are due to the super-light modes in the brane world models. It may be noted that the Newtonian potential also gets modified from various different approaches. The essential to refer here include non-commutative geometry \cite{nic,gre},minimal length in quantum gravity \cite{ali}, $f(R)$ gravity \cite{noj, cap}, $\Lambda$CDM \cite{Ham20} and the entropic force \cite{maj}. Work has been done to see the impact of modified potentials on thermodynamics of galaxy clustering \cite{4,Hameeda2,Hameeda}.\\
The effect of the cosmological constant on the clustering of galaxies has been used deriving the gravitational partition function for galaxies in a universe with a cosmological constant and the influence on distribution of galaxies has been studied \cite{1,1b}. It is important to mention the phenomenological Tohline-Kuhn modified gravity approach to the problem of dark matter \cite{kuh1,toh1,toh2}.\\
This paper is dedicated to study the clustering of galaxies by using the nonlocal aspect of gravity which is actually the nonlocal extension of general relativity. This is done in the linear weak field approximation by using the modified gravitational potential for a point mass \cite{toh1}. The work in this paper includes the Boltzmann statistical approach and Tsallis statistical approach to exclusively get the partition function and the thermodynamic properties of gravitating gas.

\section{Partition function of an interacting system in the non-local gravity using Boltzmann statistics}

 The general partition function of a system of $N$ particles of mass $m$ interacting through the modified gravitational potential
with potential energy is $\Phi$,  can be written as
\begin{eqnarray}
\label{eq2.1}
{\cal Z}_\nu&=& \frac{1}{N!}\int d^{\nu}pd^{\nu}r
\times  \exp\biggl\{-\biggl[\sum_{i=1}^{N}\frac{p_{i}^2}{2m}+\Phi(r_{1}, r_{2}, r_{3},
\dots, r_{N})\biggr] T^{-1}\biggr\},
\end{eqnarray}
The nonlocal potential energy for point mass is written as \cite{toh1}
\begin{eqnarray}
\label{eq2.2}
(\Phi_{i,j})_{nl}=-\frac{Gm^2}{r_{ij}}+\frac{Gm^2}{\lambda}\ln\bigl(\frac{r_{ij}}{\lambda}\bigr)
\end{eqnarray}
For large $N$ number of galaxies one can write
\begin{eqnarray}
\label{eq2.3}
(\Phi_{i,j})_{nl}=-\frac{N(N-1)Gm^2}{2r_{ij}}+\frac{N(N-1)Gm^2}{2\lambda}\ln\bigl(\frac{r_{ij}}{\lambda}\bigr)
\end{eqnarray}
The partition function ${\cal Z}$ in $\nu$ dimensions.
\begin{equation}
\label{eq2.4}
{\cal Z}_\nu=-\frac{1}{N!}\int\limits_{-\infty}^{\infty}d^\nu x\int\limits_{-\infty}^{\infty}d^\nu p
\exp\left[\beta\left(\frac {N(N-1)Gm^2} {2r}-\frac{N(N-1)Gm^2}{2\lambda}\ln\bigl(\frac{r}{\lambda}\bigr)-\frac {Np^2} {2m}\right)\right]
\end{equation}
Or equivalently:
\[{\cal Z}_\nu=-\frac{1}{\lambda^{-\frac{\beta N(N-1)Gm^2}{2\lambda}}N!}\left[\frac {2\pi^{\frac {\nu} {2}}} {\Gamma\left(\frac {\nu} {2}\right)}\right]^2
\int\limits_0^{\infty}r^{\nu-\frac{\beta N(N-1)Gm^2}{2\lambda}-1}\]
\begin{equation}
\label{eq2.5}
\exp\left[\beta\left(\frac {N(N-1)Gm^2} {2r}\right) \int\limits_0^{\infty}p^{\nu-1} dp\exp\left(-\frac {Np^2} {2m}\right)\right]
\end{equation}
%where $s=\frac{\beta N(N-1)Gm^2}{2\lambda}$\\

\begin{figure}[h!]
 \begin{center}$
 \begin{array}{cccc}
\includegraphics[width=60 mm]{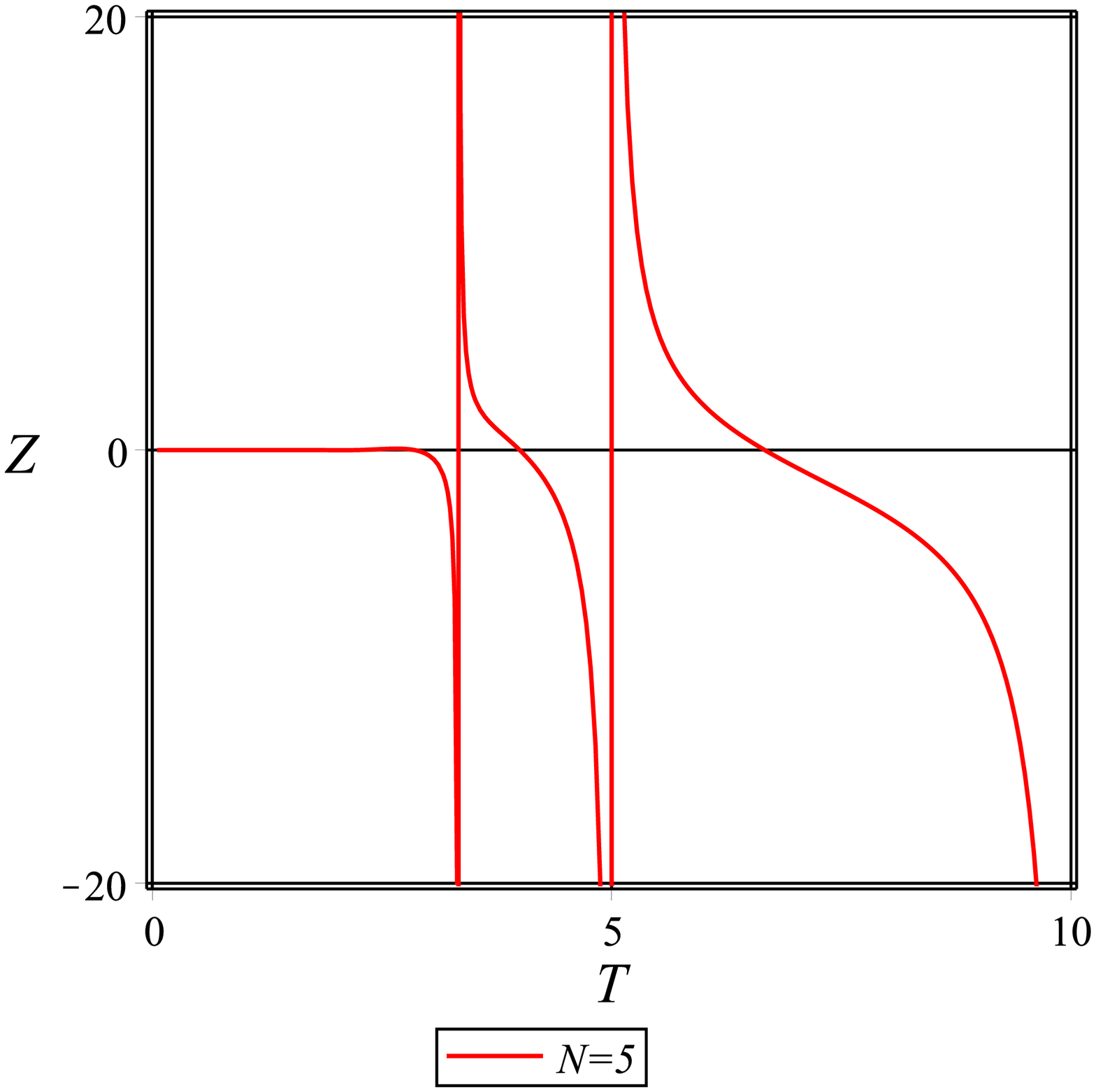}\includegraphics[width=60 mm]{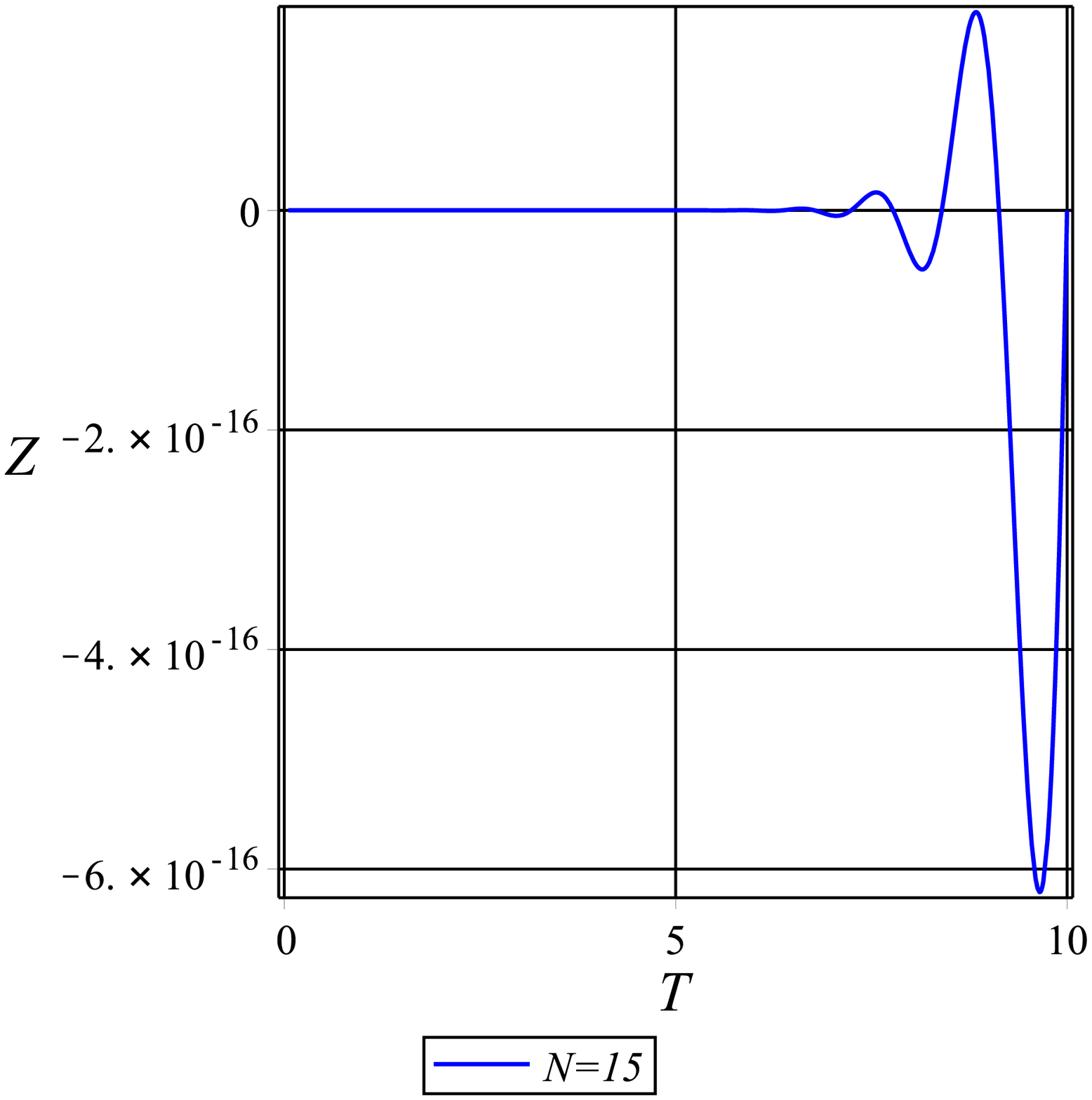}
 \end{array}$
 \end{center}
\caption{Partition function in terms of temperature with $G=M=k=1$, and $\lambda=3$.}
 \label{fig1}
\end{figure}

Using the integrals \cite{di14,gr}:
\[\int\limits_0^{\infty}r^{\nu-\frac{\beta N(N-1)Gm^2}{2\lambda}-1}dr
\exp\left[\beta\left(\frac {N(N-1)Gm^2} {2r}\right)\right]=\]
\begin{equation}
\label{eq2.6}
\cos(\pi(\nu-\frac{\beta N(N-1)Gm^2}{2\lambda}))\left(\frac{N(N-1)\beta Gm^2}{2}\right)^{\nu-\frac{\beta N(N-1)Gm^2}{2\lambda}}\Gamma\left(\frac{\beta N(N-1)Gm^2}{2\lambda}-\nu\right)
\end{equation}
and
\begin{equation}
\label{eq2.7}
\int\limits_0^{\infty}p^{\nu-1}dr\exp\left(-\beta\left(\frac {Np^2} {2m}\right)\right)=\frac{2^{-\nu} \left(\frac{N\beta}{2m}\right)^\frac{-\nu}{2}\sqrt\pi\Gamma(\nu)}{\Gamma\left(\frac{\nu+1}{2}\right)}
\end{equation}
Thus the partition function ${\cal Z}$
\[{\cal Z}=-\frac{1}{\lambda^{-s}N!}\left[\frac {2\pi^{\frac {\nu} {2}}} {\Gamma\left(\frac {\nu} {2}\right)}\right]^2
\cos(\pi(\nu-\frac{\beta N(N-1)Gm^2}{2\lambda}))\left(\frac{N(N-1)\beta Gm^2}{2}\right)^{\nu-s}\times\]
\begin{equation}
\label{eq2.8}
\Gamma\left(\frac{\beta N(N-1)Gm^2}{2\lambda}-\nu\right)\\
\frac{2^{-\nu} \left(\frac{N\beta}{2m}\right)^\frac{-\nu}{2}\sqrt\pi\Gamma(\nu)}{\Gamma\left(\frac{\nu+1}{2}\right)}
\end{equation}
For three dimensions $\nu=3$ and the partition function becomes
\[{\cal Z}=-\frac{4\lambda^{\frac{\beta N(N-1)Gm^2}{2\lambda}}\pi^{\frac{3}{2}}}{N!}\cos\pi(3-\frac{\beta N(N-1)Gm^2}{2\lambda})\left(\frac{N(N-1)Gm^2}{2}\right)^{3-\frac{\beta N(N-1)Gm^2}{2\lambda}}\left(\frac{N}{2m}\right)^{-\frac{3}{2}}\times\]
\begin{equation}
\label{eq2.9}
\beta^{\frac{3}{2}-\frac{\beta N(N-1)Gm^2}{2\lambda}}\Gamma\left(\frac{\beta N(N-1)Gm^2}{2\lambda}-3\right)
\end{equation}
Here we see that there are no poles in the partition function obtained from Boltzmann statistics provided $\frac{\beta N(N-1)Gm^2}{2\lambda}>3$. The value of $\frac{\beta N(N-1)Gm^2}{2\lambda}$ for a particular system depends on the value of non local parameter $\lambda$, which is small enough to make $\frac{\beta N(N-1)Gm^2}{2\lambda}$ greater than $3$. Thus the partition function reveals that the non local gravity provides us a divergence free gravity model. Typical behavior of the partition function illustrated by plots of Fig. \ref{fig1}.Note that the oscillation of the system is more pronounced near the origin of the temperatures. The permitted physical temperatures are those for which ${\cal Z}>0$.

\section{Internal energy and the specific heat of the system in nonlocal gravity using Boltzmann partition function}

\begin{figure}[h!]
 \begin{center}$
 \begin{array}{cccc}
\includegraphics[width=60 mm]{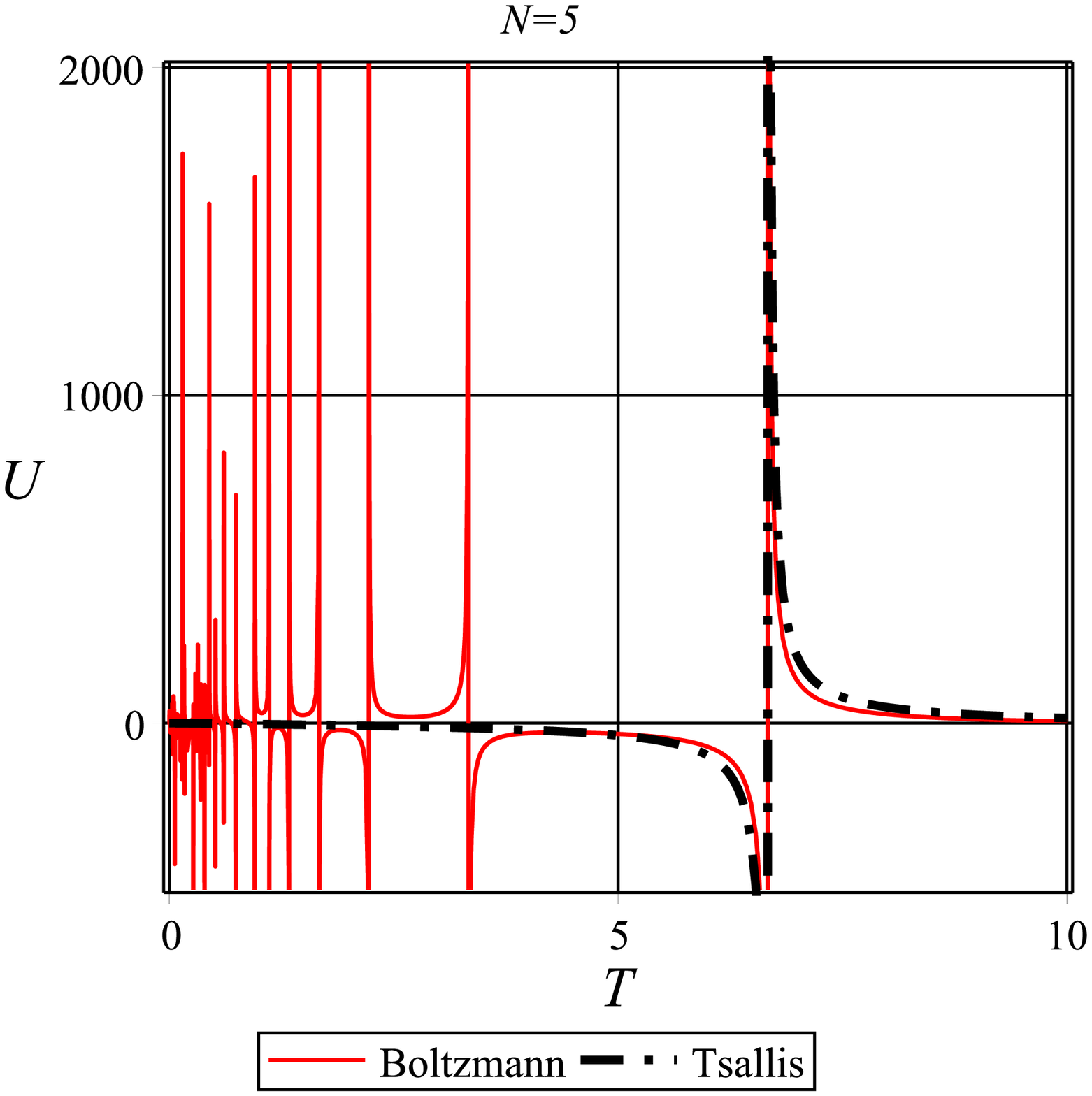}\includegraphics[width=60 mm]{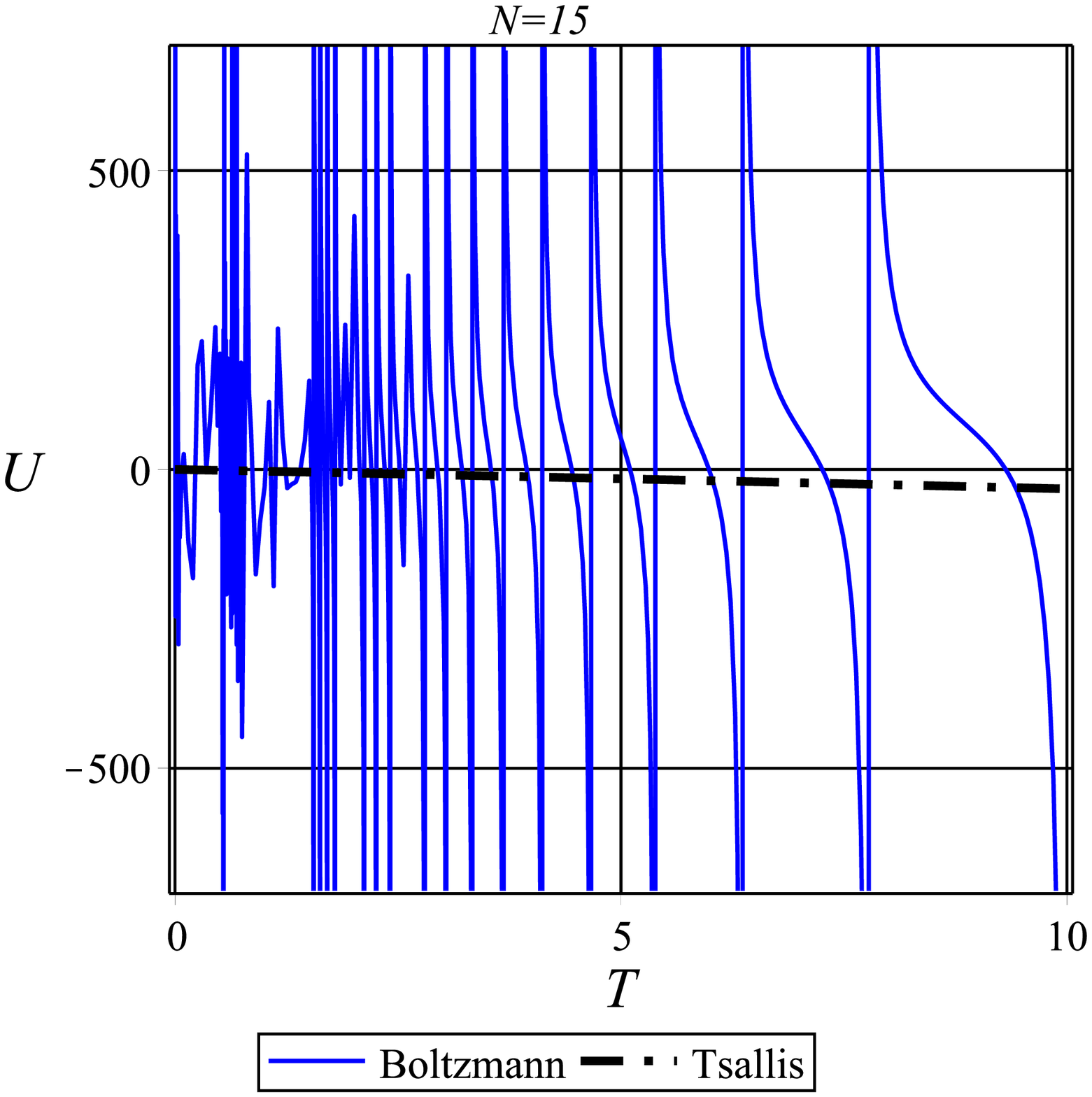}
 \end{array}$
 \end{center}
\caption{Internal energy in terms of temperature with $G=M=k=1$, and $\lambda=3$.}
 \label{fig2}
\end{figure}

 Internal energy $U$ is obtained as
\begin{equation}
\label{eq3.1}
U=-\frac{1}{{\cal Z}}\frac{\partial {\cal Z}}{\partial\beta}
\end{equation}
\[U=\left[-\frac{3}{2\beta}+\frac{N(N-1)Gm^2}{2\lambda}\left(1+\ln \frac{\beta N(N-1)
Gm^2}{2\lambda}-\pi\tan\pi\left(3-\frac{\beta N(N-1)Gm^2}{2\lambda}\right)-\right.\right.\]
\begin{equation}
\label{eq3.2}
\left.\left.\psi\left(\frac{\beta N(N-1)Gm^2}{2\lambda}-3\right)\right)\right]
\end{equation}
In the plot of the Fig. \ref{fig2} we can see periodic nature of the internal energy.

 Specific heat $C_v$ is:
\[C_V=\left[-\frac{3k}{2}-{k\frac{\beta N(N-1)Gm^2}{2\lambda}}\left(1+\right.\right.\]
\[\pi \frac{\beta N(N-1)Gm^2}{2\lambda}\sec^2\pi\left(3-\frac{\beta N(N-1)Gm^2}{2\lambda}\right)-\]
\begin{equation}
\label{eq3.3}
\left.\left.\frac{\beta N(N-1)Gm^2}{2\lambda}\psi^{\prime}\left(\frac{\beta N(N-1)Gm^2}{2\lambda}-3\right)\right)\right]
\end{equation}

\begin{figure}[h!]
 \begin{center}$
 \begin{array}{cccc}
\includegraphics[width=70 mm]{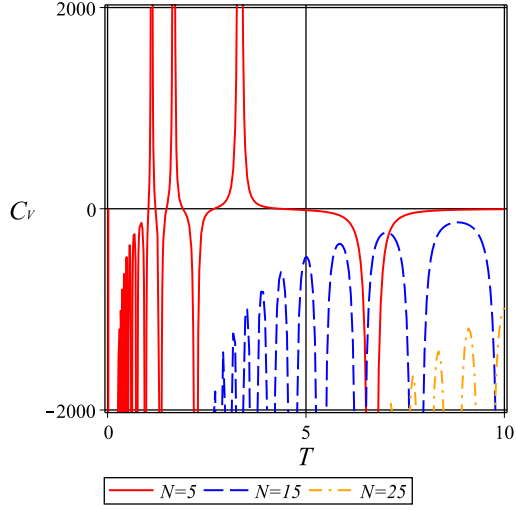}
 \end{array}$
 \end{center}
\caption{Specific heat in terms of temperature with $G=M=k=1$, and $\lambda=3$.}
 \label{fig3}
\end{figure}

 Larger $N$ yields to the negative specific heat which means that the system is self-gravitating  (see Fig. \ref{fig3}).

\section{Thermodynamic properties of a system in nonlocal gravity using Boltzmann statistics}
\begin{figure}[h!]
 \begin{center}$
 \begin{array}{cccc}
\includegraphics[width=70 mm]{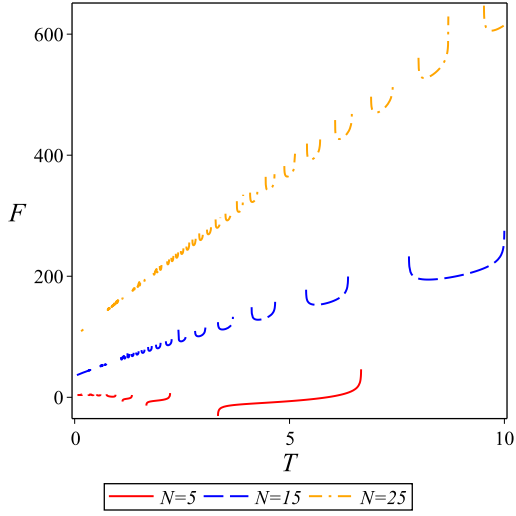}
 \end{array}$
 \end{center}
\caption{Helmholtz free energy in terms of temperature with $G=M=k=1$, and $\lambda=3$.}
 \label{fig4}
\end{figure}

Helmhotz free energy $F$ is obtained using
\begin{equation}
\label{eq4.1}
F=-\frac{1}{\beta}\ln {\cal Z}
\end{equation}
Thus $F$ is obtained and written as
\[F=-\frac{1}{\beta}\ln\left[\frac{4\lambda^{\frac{\beta N(N-1)Gm^2}{2\lambda}}\pi^{\frac{3}{2}}}{N!}\cos\pi(3-\frac{\beta N(N-1)Gm^2}{2\lambda})\times\right.\]
\begin{equation}
\label{eq4.2}
\left.\left(\frac{N(N-1)Gm^2}{2}\right)^{3-\frac{\beta N(N-1)Gm^2}{2\lambda}}\left(\frac{N}{2m}\right)^{-\frac{3}{2}}\beta^{\frac{3}{2}-\frac{\beta N(N-1)Gm^2}{2\lambda}}\Gamma\left(\frac{\beta N(N-1)Gm^2}{2\lambda}-3\right)\right]
\end{equation}
In Fig. \ref{fig4} we can see typical behavior of Helmholtz free energy.
The entropy $S$ is obtained from
\begin{equation}
\label{eq4.3}
%\label{eq4.4}
TS=U-F
\end{equation}
Thus:
\[ST=\left[-\frac{3}{2\beta}+\frac{s}{\beta}\left(1+\ln \frac{\beta N(N-1)Gm^2}{2\lambda}-\pi\tan\pi\left(3-\frac{\beta N(N-1)Gm^2}{2\lambda}\right)-\right.\right.\]
\[\left.\left.\psi(\frac{\beta N(N-1)Gm^2}{2\lambda}-3)\right)\right]
+\frac{1}{\beta}\ln\left[\frac{4\lambda^{\frac{\beta N(N-1)Gm^2}{2\lambda}}\pi^{\frac{3}{2}}}{N!}\cos\pi\left(3-\frac{\beta N(N-1)Gm^2}{2\lambda}\right)\right.\]
\begin{equation}
\label{eq4.4}
%\label{eq4.5}
\left.\left(\frac{N(N-1)Gm^2}{2}\right)^{3-\frac{\beta N(N-1)Gm^2}{2\lambda}}\left(\frac{N}{2m}\right)^{-\frac{3}{2}}\beta^{\frac{3}{2}-\frac{\beta N(N-1)Gm^2}{2\lambda}}\Gamma\left(\frac{\beta N(N-1)Gm^2}{2\lambda}-3\right)\right]
\end{equation}
Pressure $P$ is obtained using
\begin{equation}
\label{eq4.5}
PV=\frac{2N}{3}U
\end{equation}
Thus:
\[PV=\frac{2N}{3}\left[-\frac{3}{2\beta}+\frac{\frac{\beta N(N-1)Gm^2}{2\lambda}}{\beta}\left(1+\ln \frac{\beta N(N-1)Gm^2}{2\lambda}-\pi\tan\pi\left(3-\frac{\beta N(N-1)Gm^2}{2\lambda}\right)\right.\right.-\]
\begin{equation}
\label{eq4.6}
\left.\left.\psi\left(\frac{\beta N(N-1)Gm^2}{2\lambda}-3\right)\right)\right]
\end{equation}
and:
\[\beta PV=\frac{2N}{3}\left[-\frac{3}{2}+\frac{\beta N(N-1)Gm^2}{2\lambda}\left(1+\ln \frac{\beta N(N-1)Gm^2}{2\lambda}-\pi\tan\pi\left(3-\frac{\beta N(N-1)Gm^2}{2\lambda}\right)\right.\right.\]
\begin{equation}
\label{eq4.7}
\left.\left.-\psi\left(\frac{\beta N(N-1)Gm^2}{2\lambda}-3\right)\right)\right]
\end{equation}
The chemical potential $\mu$ is obtained using
\begin{equation}
\label{eq4.8}
\mu=\left(\frac{\partial F}{\partial N}\right)_T
\end{equation}
Thus:
\[\mu=\frac{1}{\beta}\left[\frac{3}{2N}+\ln N+\frac{(2N-1)\frac{\beta N(N-1)Gm^2}{2\lambda}}{N(N-1)}\left(1+\ln \frac{\beta N(N-1)Gm^2}{2\lambda}-\right.\right.\]
\begin{equation}
\label{eq4.9}
\left.\left.\pi\tan\pi\left(3-\frac{\beta N(N-1)Gm^2}{2\lambda}\right)-\frac{3}{\frac{\beta N(N-1)Gm^2}{2\lambda}}-\psi\left(\frac{\beta N(N-1)Gm^2}{2\lambda}-3\right)\right)\right]
\end{equation}
and
\[{\beta N\mu}=\left[\frac{3}{2}+N\ln N+\frac{(2N-1)\frac{\beta N(N-1)Gm^2}{2\lambda}}{(N-1)}\left(1+\ln \frac{\beta N(N-1)Gm^2}{2\lambda}\right.\right.-\]
\begin{equation}
\label{eq4.10}
\left.\left.\pi\tan\pi\left(3-\frac{\beta N(N-1)Gm^2}{2\lambda}\right)-\frac{3}{\frac{\beta N(N-1)Gm^2}{2\lambda}}-\psi\left(\frac{\beta N(N-1)Gm^2}{2\lambda}-3\right)\right)\right]
\end{equation}

\begin{eqnarray}
\label{eq4.8}
\mu &=&\frac{3}{2N\beta}+\frac{\ln}{\beta} N+\frac{(2N-1)s}{\beta N(N-1)}\nonumber\\
&\times&\Big(1+\ln s-\pi\tan\pi(3-s) -\frac{3}{s}  -\psi(s-3)\Big).
\end{eqnarray}

\begin{figure}[h!]
 \begin{center}$
 \begin{array}{cccc}
\includegraphics[width=70 mm]{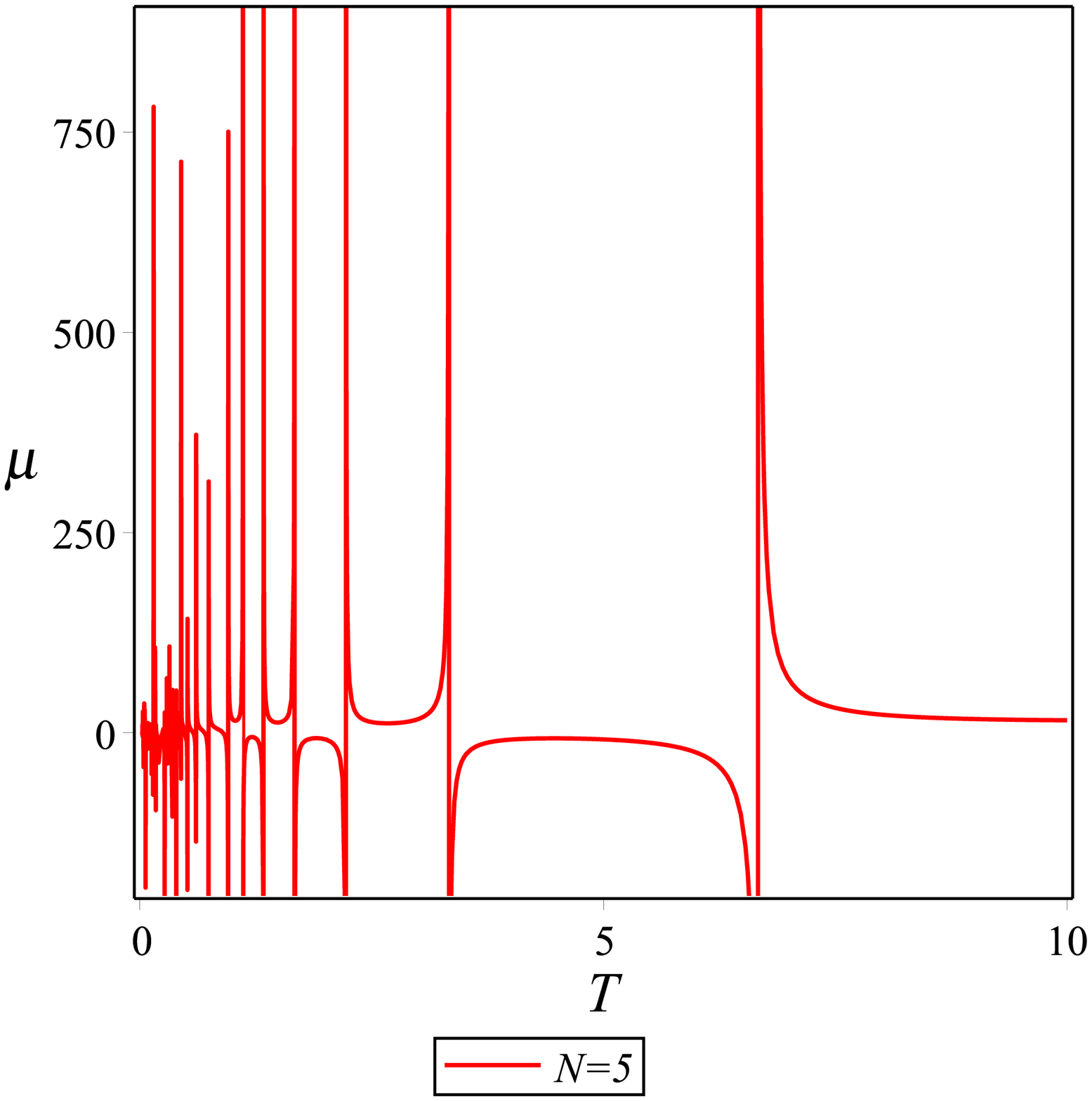}\includegraphics[width=70 mm]{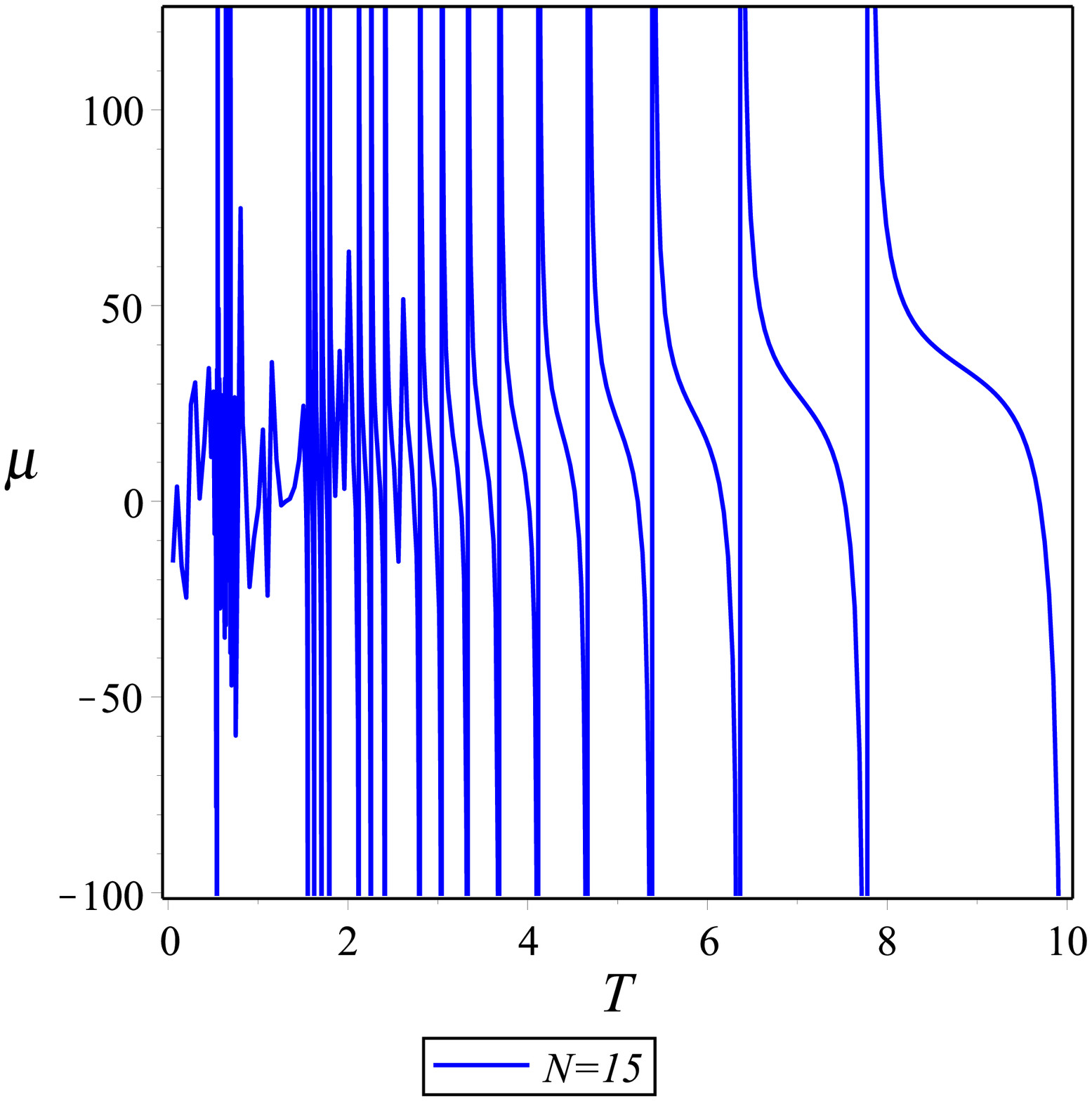}
 \end{array}$
 \end{center}
\caption{Chemical potential in terms of temperature with $G=M=k=1$, and $\lambda=3$.}
 \label{fig5}
\end{figure}

\section{Distribution function from Boltzmann statistics}

 The grand canonical partition function $Z_G$ can be defined as
\begin{equation}
\label{eq5.1}
\ln Z_G=\beta PV
\end{equation}

 From chemical potential $\mu$ we find the fugacity $z$ as
\begin{equation}
\label{eq5.2}
z^N=e^{N\beta\mu}
\end{equation}
The distribution function follows as
\begin{equation}
\label{eq5.3}
F(N)=\frac{z^N{\cal Z}}{Z_G}
\end{equation}
The distribution function follows as
\[F(N)=\frac{4\lambda^{\frac{\beta N(N-1)Gm^2}{2\lambda}}\pi^{\frac{3}{2}}N^N}{N!}\cos\pi\left(3-\frac{\beta N(N-1)Gm^2}{2\lambda}\right)\left(\frac{N(N-1)Gm^2}{2}\right)^{3-\frac{\beta N(N-1)Gm^2}{2\lambda}}\]
\[\left(\frac{N}{2m}\right)^{-\frac{3}{2}}\beta^{\frac{3}{2}-\frac{\beta N(N-1)Gm^2}{2\lambda}}\Gamma(s-3)\]
\[\exp\biggl{\{}\frac{3}{2}+\frac{(2N-1)s}{(N-1)}\left(1+\ln \frac{\beta N(N-1)Gm^2}{2\lambda}-\pi\tan\pi\left(3-\frac{\beta N(N-1)Gm^2}{2\lambda}\right)-\right.\]
\[\left.\frac{3}{\frac{\beta N(N-1)Gm^2}{2\lambda}}-\psi\left(\frac{\beta N(N-1)Gm^2}{2\lambda}-3\right)\right)\]
\[-\frac{2N}{3}\left(-\frac{3}{2}+\frac{\beta N(N-1)Gm^2}{2\lambda}\left(1+\ln \frac{\beta N(N-1)Gm^2}{2\lambda}-\pi\tan\pi\left(3-\frac{\beta N(N-1)Gm^2}{2\lambda}\right)-\right.\right.\]
\begin{equation}
\label{eq5.4}
\left.\left.\psi\left(\frac{\beta N(N-1)Gm^2}{2\lambda}-3\right)\right)\right)\biggr{\}}
\end{equation}

Note that the permitted physical temperatures are those for which $F(N)>0$.

\begin{figure}[h!]
 \begin{center}$
 \begin{array}{cccc}
\includegraphics[width=60 mm]{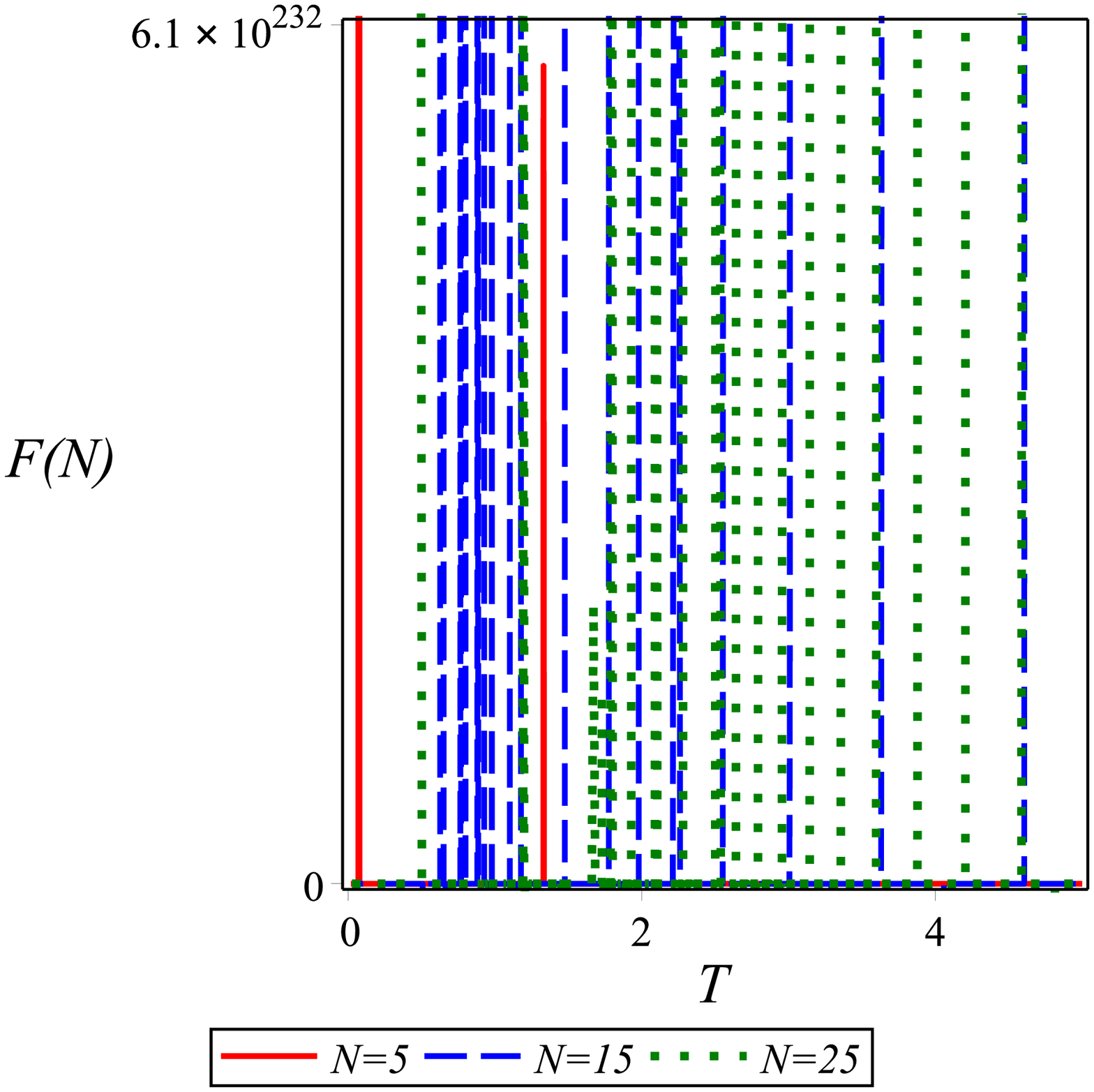}\includegraphics[width=60 mm]{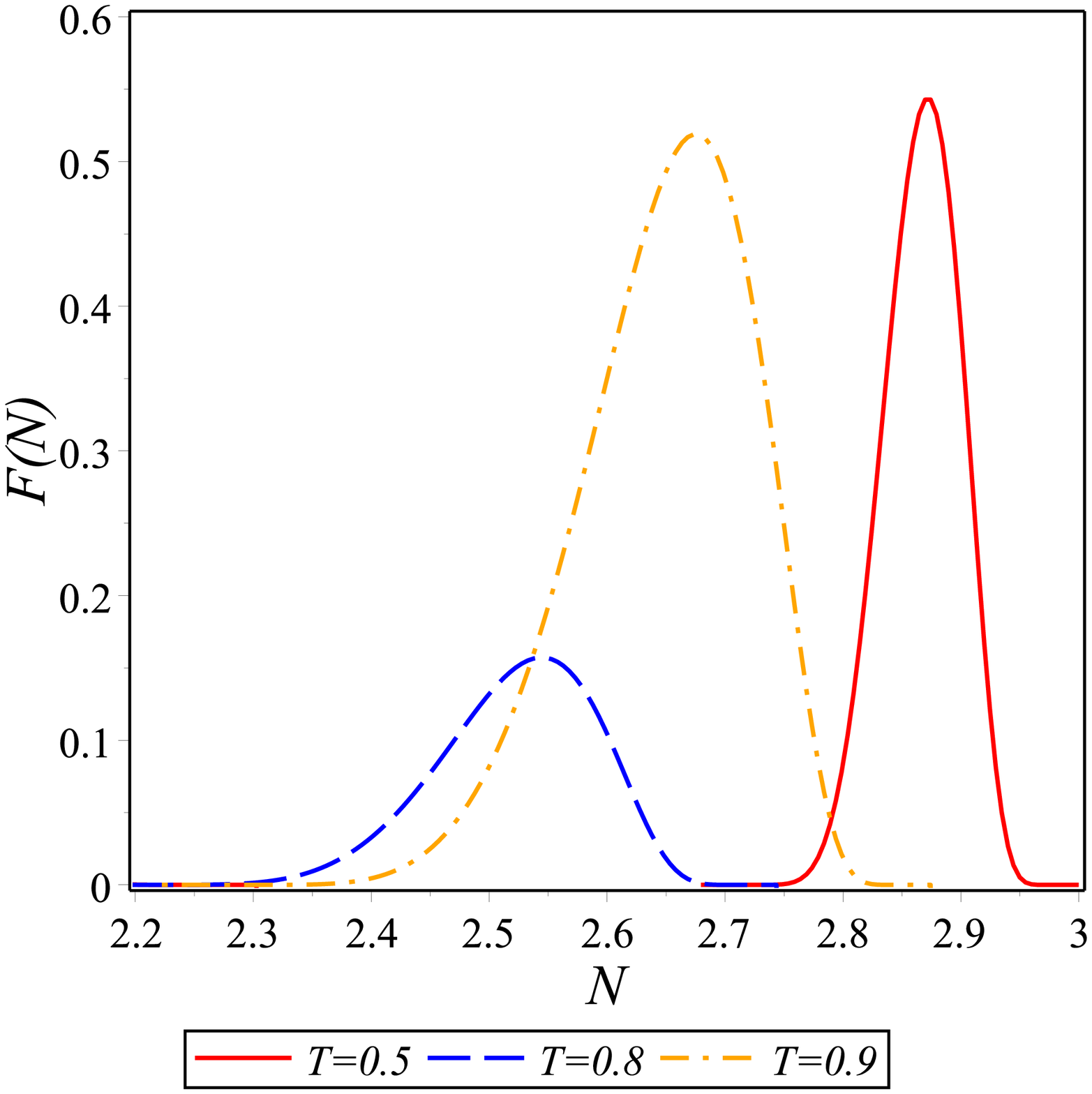}
 \end{array}$
 \end{center}
\caption{Distribution function in terms of temperature with $G=M=k=1$, and $\lambda=3$.}
 \label{fig6}
\end{figure}

The temperature profile for the sample of ten clusters of galaxies observed by Chandra telescope has been tabulated \cite{obser,rav}. The range is from $2.26$ keV to $19.34$ keV for ten clusters. The distribution function as plotted in Fig. \ref{fig6} in this observational range seems to show more positive peaks, thus in agreement with the observational data available to us from the Chandra Data.

\section{Tsallis statistical analysis of the non-local gravitational interactions}

 We repeat the whole analysis using Tsallis statistics.

 Tsallis q-exponential is defined as the distribution:
\begin{equation}
\label{eq6.1}
e_q(x)=[1+(q-1)x]^{\frac{1}{q-1}}
\end{equation}
Or equivalently
\begin{equation}
\label{ep6.2}
e_q(x)=
\begin{cases}
[1+(q-1)x]^{\frac{1}{q-1}}\;\;\;;\;\;\;1+(q-1)x>0\\
0\;\;\;;\;\;\;1+(q-1)x<0
\end{cases}
\end{equation}

\section{Tsallis partition function for system in non local gravity}

 Let's consider the distribution
$\frac {1} {r}=PV\frac {1} {r}$.
Then $\frac {1} {r}\mid_{r=0}=0$.
First we do the calculation in $\nu$ dimensions. Let  $q>1$.
The nonlocal potential energy for point mass is written as
\begin{eqnarray}
\label{eq7.1}
%\label{eq2.3}
(\Phi_{i,j})_{nl}=-\frac{Gm^2}{r_{ij}}+\frac{Gm^2}{\lambda}\ln\bigl(\frac{r_{ij}}{\lambda}\bigr)
\end{eqnarray}
For large $N$ number of galaxies one can write
\begin{eqnarray}
\label{eq7.2}
%\label{eq2.4}
(\Phi_{i,j})_{nl}=-\frac{N(N-1)Gm^2}{2r_{ij}}+\frac{N(N-1)Gm^2}{2\lambda}\ln\bigl(\frac{r_{ij}}{\lambda}\bigr)
\end{eqnarray}
Keeping in view the complicated procedure of solving integrals in Tsallis statistics we use $\ln(1+z)=z$ approximation $\ln(\frac{\lambda}{r}-1+1)=\frac{\lambda}{r}-1$
\begin{eqnarray}
\label{eq7.3}
%\label{eq2.5}
(\Phi_{i,j})_{nl}=-2\frac{N(N-1)Gm^2}{2r_{ij}}+\frac{N(N-1)Gm^2}{2\lambda}
\end{eqnarray}
For the partition we have
\begin{equation}
\label{eq7.4}
%\label{eq2.6}
{\cal Z}_\nu=\int\limits_{-\infty}^{\infty}d^\nu x\int\limits_{-\infty}^{\infty}d^\nu p
\left[1+(q-1)\beta\left(\frac {N(N-1)Gm^2} {r}-\frac{N(N-1)Gm^2}{2\lambda}-\frac {Np^2} {2m}\right)\right]_+^{\frac {1} {q-1}}
\end{equation}
And evaluating the angular integral we obtain:
\begin{equation}
\label{eq7.5}
%\label{eq2.7}
{\cal Z}_\nu=\left[\frac {2\pi^{\frac {\nu} {2}}} {\Gamma\left(\frac {\nu} {2}\right)}\right]^2
\int\limits_0^{\infty}r^{\nu-1}dr\int\limits_0^{\infty}p^{\nu-1} dp
\left[1+(q-1)\beta\left(\frac {N(N-1)Gm^2} {r}-\frac{N(N-1)Gm^2}{2\lambda}-\frac {Np^2} {2m}\right)\right]_+^{\frac {1} {q-1}}
\end{equation}
Taking into account that:
\begin{equation}
\label{eq7.6}
%\label{eq2.8}
1+(q-1)\beta\left(\frac {N(N-1)Gm^2} {r}-\frac{N(N-1)Gm^2}{2\lambda}-\frac {Np^2} {2m}\right)>0
\end{equation}
should be:
\begin{equation}
\label{eq7.7}
%\label{eq2.9}
r<\frac {2(q-1)\beta N(N-1)Gm^3\lambda}
{\beta N(N-1)(q-1)Gm^3+\beta N(q-1)p^2\lambda-2m\lambda}
\end{equation}
We have then:
\[{\cal Z}_\nu=\left[\frac {2\pi^{\frac {\nu} {2}}} {\Gamma\left(\frac {\nu} {2}\right)}\right]^2
\int\limits_0^{\infty}p^{\nu-1}dp
\int\limits_0^{r_0} r^{\nu-1} dr\times\]
\begin{equation}
\label{eq7.8}
%\label{eq2.10}
\left[1+(q-1)\beta\left(\frac {N(N-1)Gm^2}{r}-\frac{N(N-1)Gm^2}{2\lambda}-\frac {Np^2} {2m}\right)\right]^{\frac {1} {q-1}}
\end{equation}
Evaluating the integral we arrive to the result:
\[Z_\nu=\frac{1}{2}\left[\frac{2\pi^\frac{\nu} {2}}{\Gamma\left(\frac {\nu} {2}\right)}\right]^2
(2m\lambda)^{-\frac {1} {q-1}}[2(q-1)\beta N(N-1)Gm^3\lambda)^{\nu}\]
\[[\beta N(q-1)\lambda)^{-\frac {\nu} {2}}
[\beta N(N-1)(q-1)Gm^3-2m\lambda]^{\frac {3} {2}}\]
\begin{equation}
\label{eq7.9}
%\label{eq2.11}
B\left(1+\frac {1} {q-1},\nu-\frac {1} {q-1}\right)
B\left(\frac {\nu} {2},\frac {\nu} {2}-\frac {1} {q-1}\right)
\end{equation}
After simplification we get the partition function in three dimensions as
\begin{equation}
\label{eq7.10}
%\label{eq2.12}
Z=\frac {16} {27}\pi^3\frac {(\beta N\lambda)^{-\frac {3} {2}}} {(m\lambda)^3}
[\beta N(N-1)Gm^3\lambda]^3
[2\beta N(N-1)Gm^3-2m\lambda]^{\frac {3} {2}}
\end{equation}
Finally  we have:
\begin{equation}
\label{eq7.11}
%\label{eq2.12}
Z=\frac {2^{\frac {11} {2}}\pi^3} {27}
\frac {(\beta N\lambda)^{-\frac {3} {2}}} {(m\lambda)^3}
[\beta N(N-1)Gm^3\lambda]^3
[\beta N(N-1)Gm^3-3m\lambda]^{\frac {3} {2}}
\end{equation}
In Tsallis statistical analysis of the non local gravity we again find no poles in the partition function and thus we again emphasis on the fact that the non local gravity is a divergence free model of gravity.

\section{Tsallis internal energy and specific heat of an interacting system in non local gravity}

 We calculate the internal energy and specific heat to see the other important properties of the system. We use:
\[{\cal Z}_\nu<{\cal U}>_\nu=\left[\frac {2\pi^{\frac {\nu} {2}}} {\Gamma\left(\frac {\nu} {2}\right)}\right]^2
\int\limits_0^{\infty}p^{\nu-1}dp
\int\limits_0^{r_0} r^{\nu-1} dr\times\]
\[\left(-\frac {N(N-1)Gm^2}{r}+\frac{N(N-1)Gm^2}{2\lambda}+\frac {Np^2} {2m}\right)\]
\begin{equation}
\label{eq8.1}
%\label{eq2.10}
\left[1+(q-1)\beta\left(\frac {N(N-1)Gm^2}{r}-\frac{N(N-1)Gm^2}{2\lambda}-\frac {Np^2} {2m}\right)\right]^{\frac {1} {q-1}}
\end{equation}
Then, proceeding in the same way as for the partition function, we obtain:
\[<{\cal U}>=\frac {N(N-1)Gm^2} {2\lambda}+\frac {1} {{\cal Z}}\left\{\frac {2^{\frac {13} {2}}\pi^3} {135}
\frac {N(N-1)Gm^2} {(m\lambda)^3}(\beta N\lambda)^{-\frac {3} {2}}\times\right.\]
\[\left[\beta N(N-1)Gm^3\lambda\right]^2\left[\beta N(N-1)Gm^3-3m\lambda\right]^{\frac {5} {2}}-
\frac {2^{\frac {9} {2}}\pi^3} {135}\frac {N} {m}\times\]
\begin{equation}
\label{eq8.2}
\left.\frac {(\beta N\lambda)^{-\frac {5} {2}}} {(m\lambda)^3}
\left[\beta N(N-1)Gm^3\lambda\right]^3\left[\beta N(N-1)Gm^3-3m\lambda\right]^{\frac {5} {2}}\right\}
\end{equation}

\begin{equation}
\label{8.3}
<{\cal U}>=\frac {3} {10}\left(\frac {N(N-1)Gm^2} {\lambda}-3kT\right)
\end{equation}
See black dash dotted line of Fig. \ref{fig2}.
\begin{equation}
\label{eq8.4}
C_v=-\frac{9k}{10}
\end{equation}
Again the specific heat corresponds to a self-gravitating system. However, unlike our Boltzmann statistic, an oscillating system, in this case the system does not oscillate.
Note that we have found a new phenomenon that we have not yet been able to explain convincingly. The specific heat of the system is independent of $ N $. The only explanation we have found so far is that for this system, if we add more objects to it, the specific heat corresponding to each of the objects increases in such a way that the specific heat of the system remains constant. We leave the treatment of this interesting problem to the consideration of the scientific community.

\section{Equations of State}

 Other thermodynamic properties are calculated using Tsallis statistics are as under.
Then, we can evaluate the entropy ${\cal S}$ with the formula
\begin{equation}
\label{eq9.1}
%\label{eq5.1}
{\cal S}=\ln_{\frac {4} {3}}{\cal Z}+{\cal Z}^{-\frac {1} {3}}\beta<{\cal U}>
\end{equation}
It can be rewritten in the form:
\begin{equation}
\label{eq9.2}
%\label{eq5.2}
{\cal S}={-3({\cal Z}^\frac{-1}{3}-1)}+{\cal Z}^{-\frac {1} {3}}\beta<{\cal U}>
\end{equation}
or equivalently:
\begin{equation}
\label{eq9.3}
%\label{eq5.3}
{\cal S}=3+(\beta<{\cal U}>-3){\cal Z}^{-\frac{1}{3}}
\end{equation}
Thus:
\[{\cal S}=3-\left[\frac{3\beta} {10}\left(3kT-\frac {N(N-1)Gm^2} {\lambda}\right)+3\right]\times\]
\begin{equation}
\label{eq9.4}
\left[\frac {16} {27}
[2\beta N(N-1)^2G^2m^5]^{\frac {3} {2}}(2\frac{\beta N(N-1)Gm^2}{2\lambda}-1)^{\frac {3} {2}}\right]^{-\frac{1}{3}}
\end{equation}

 The free Helmholtz energy is given by:
\begin{equation}
\label{eq9.5}
%\label{eq5.5}
{\cal F}=<{\cal U}>-T{\cal S}
\end{equation}
Thus:
\[{\cal F}=\frac {3} {10}\left(\frac {N(N-1)Gm^2} {\lambda}-3kT\right)+T
\left[\frac{3\beta} {10}\left(3kT-\frac {N(N-1)Gm^2} {\lambda}\right)+3\right]\times\]
\begin{equation}
\label{eq9.6}
\left[\frac {16} {27}[2\beta N(N-1)^2G^2m^5]^{\frac {3} {2}}(2\frac{\beta N(N-1)Gm^2}{2\lambda}-1)^{\frac {3} {2}}\right]^{-\frac{1}{3}}-3T
\end{equation}
And the pressure $P$ as:
\begin{equation}
\label{eq9.7}
P=\frac {N} {V}\frac {2<{\cal U}>} {3N-2}
\end{equation}
Then:
\begin{equation}
\label{eq9.8}
P=\frac {N} {V}\frac {6} {10(3N-2)}
\left(\frac {N(N-1)Gm^2} {\lambda}-3kT\right)
\end{equation}

The parameter $\lambda$ which is the basic nonlocality length scale whose value has been discussed to be $3\pm 2kpc$
\cite{rah}. For the $\lambda\to\infty$ the nonlocality as well as the effective dark matter disappear. It has been shown that for galaxies  the ratio of baryonic diameter to dark matter must be above the fixed nonlocal scale length \cite{bah}.

\section{Conclusion}
 It was well known that the Boltzmann partition function of gravity could not be calculated because the integral that defines it is exponentially divergent. In 2018 this problem was solved by Plastino and Rocca \cite{di14, jpco, xz16} by using the generalization of the dimensional regularization of Bollini and Giambiagi \cite{xz12, xz14, prd}. This generalization was based on the general quantification method of QFT's \cite{la12, la14, la16, la18} using Ultradistributions of Sebastiao
e Silva, also known as Ultrahyperfunctions \cite{jss, hasumi, tp8}. In this paper we have shown that the use of such a methodology is not necessary in the case of a specific non-local theory of gravity. To calculate the partition function, in this case, it is enough with a simple analytical extension of an integral obtained by Gradshtein and Rizhick and published in their famous "Table of Integrals, Serias and Products" \cite{gr}. With this simple extension we have shown that the result of the calculation of the partition function is finite. This is a remarkable result expected by the scientific community specialized in the treatment of
non-local theories of gravity. Even more remarkable is the result of the calculation of the partition function, using the Tsallis statistics. In this case the partition function is directly finite using the Lebesgue-Stiejel integrals commonly used in statistical mechanics. We have also calculated using both statistics, the internal energy and the specific heat for this model of gravity, obtaining a finite result. Finally we have evaluated the thermodynamic functions and obtained the corresponding state functions.
The oscillatory nature of partition function and the corresponding thermodynamics can be interpreted on the same footing as oscillations of energy density explained as screening mechanism in condensed matter physics called as Friedel oscillations \cite{frie} produced when a positively charged impurity is inserted into a cold metal and the same has been dubbed for the nonlocal effect produced by a gravitational impurity in the Minkowski vacuum, which causes spatial oscillations what has been beautifully named as gravitational Friedel oscillations \cite{jens}. These oscillations make their presence in the thermodynamic properties of the system interacting in the nonlocal gravity. These fluctuations might become observable in future as has been stressed in other works\cite{acc1,acc2}. The physical significance of these oscillations is believed to be relevant, not only because they are considered as spurious effects of higher derivative theories of gravity  at the Pauli-Villars regularization scale which arise due to the presence of complex poles \cite{acc1,gia}, but also because they have been also shown to survive ghost-free limit \cite{cal,buo}. An aspect to be highlighted is the difference in the predictions of the Boltzmann and Tsallis statistics. With the first the system is oscillating and with the second it is not.

\end{document}